\documentclass{aa} 

\usepackage{graphicx}
\usepackage{txfonts}
\usepackage{natbib}
\usepackage{url}
\bibpunct{(}{)}{;}{a}{}{,}

\newcommand{\be}{\begin{equation}}
\newcommand{\ee}{\end{equation}}

\def\apjl{ApJL}
\def\apjs{ApJS}

\newcommand{\Mpc}{$h^{-1}$\thinspace Mpc}

\newcommand{\vmh}{h^{-1}\mathrm{Mpc} }

\begin{document}  

\title{Shell-like structures in our cosmic neighbourhood
} 

\author {M.~Einasto\inst{1} \and  P.~Hein\"am\"aki\inst{6}
\and L.J.~Liivam\"agi\inst{1,3} \and V.J.~Mart\'{\i}nez\inst{4}
\and L.~Hurtado-Gil\inst{4} \and P.~Arnalte-Mur\inst{4} \and P.~Nurmi\inst{6}
\and J.~Einasto\inst{1,2,5}  \and  E.~Saar\inst{1,2}
}

\institute{Tartu Observatory, 61602 T\~oravere, Estonia
\and
Estonian Academy of Sciences,  EE-10130 Tallinn, Estonia
\and
Institute of Physics, Tartu University, T\"ahe 4, 51010 Tartu, Estonia
\and 
Observatori Astron\`omic, Universitat de Val\`encia, 
c/ Catedr\`atic Jos\`e Beltr\`an, 2, E-46980 Paterna, Val\`encia, Spain
\and 
ICRANet, Piazza della Repubblica 10, 65122 Pescara, Italy
\and 
Tuorla Observatory, University of Turku, V\"ais\"al\"antie 20, Piikki\"o, Finland
}

\authorrunning{M. Einasto et al. }

\offprints{M. Einasto}

\date{ Received   / Accepted   }

\titlerunning{Shell}

\abstract
{
Signatures of the processes in the early Universe are imprinted
in the cosmic web. 
Some of them may define shell-like structures 
characterised  by typical scales. Examples of such structures are shell-like
systems of galaxies, which are interpreted as a signatures
of  the baryon acoustic oscillations. 
}
{
We search for shell-like structures in the distribution of nearby rich clusters
of galaxies drawn from the SDSS DR8.
}
{
We calculated the distance distributions between rich clusters of galaxies
and groups and clusters of various richness, searched for the maxima in 
the distance distributions and selected
candidates of shell-like structures. We analysed the 
space distribution of  groups and clusters that form shell walls.
}
{We find six possible  candidates of shell-like structures, in which
galaxy clusters have the maximum in their distance distribution
to other galaxy groups and clusters at a distance of about 
$120 - 130$~\Mpc.  Another, less probable
maximum is found at a distance of about $240$~\Mpc.
The rich galaxy cluster A1795, which is
the central cluster of the Bootes supercluster,
has the highest maximum in the distance distribution of all 
other surrounding groups and clusters in our rich cluster sample. It
lies at a distance of about $120$~\Mpc. The structures of galaxy systems
that cause this maximum form an almost
complete shell of galaxy groups, clusters,
and superclusters. 
The richest systems in the nearby universe, 
the Sloan Great Wall, the Corona Borealis supercluster, and the Ursa
Major supercluster, are among them. 
The probability that we obtain maxima like this from random
distributions is lower than 0.001.
}
{ 
Our results confirm that shell-like structures  can be found 
in the distribution of nearby galaxies and their systems. The radii
of the possible shells are larger than expected for a 
baryonic acoustic oscillations (BAO) shell
($\approx109$~\Mpc\ versus $\approx120 - 130$~\Mpc),
and they are determined by very rich galaxy clusters and 
superclusters. In contrast, BAO shells are barely seen in the galaxy distribution.
We discuss possible consequences of these differences. 
}


\maketitle

\section{Introduction} 
\label{sect:intro} 

One  of  the  most  remarkable achievements  of  contemporary  cosmology  is the
discovery of the cosmic web --  a complex hierarchical network of galaxy systems
in which galaxies, galaxy groups, clusters, and superclusters form interconnected
systems that are separated by voids of various sizes
\citep{1975ATsir.895....2E, 1977TarOP...1A...1J, 
1978IAUS...79..241J, 1978ApJ...222..784G, 1979ApJ...230..648C, 1981ApJ...248L..57K,
1982Natur.300..407Z, 1983ARA&A..21..373O, 1986ApJ...302L...1D}.
A short review of the early studies of the cosmic web can be found in
 \citet{2011A&A...534A.128E}.
The properties of the cosmic web are
determined by  various processes  -- chaotic  quantum fluctuations  in the  very
early Universe, the oscillations  of the hot baryon-dark-matter-radiation plasma
before  the  recombination,  and  the growth  of  density  perturbations  that is due to
gravitational instability after the recombination. 
These processes may be reflected in 
the geometrical pattern of galaxies and
galaxy systems by characteristic scales. 

As proposed by \citet{1988Natur.334..129K}, the skeleton of the cosmic web is
fixed by processes during or just after the inflation. 
The positions of high-density
peaks and voids do not change much during the cosmic evolution, only the amplitude of
over- and underdensities grows with time
\citep[][and references therein]{1996Natur.380..603B, 2009LNP...665..291V,
2011A&A...531A.149S}.
The growth of density perturbations in the post-recombination
era is quite well studied 
\citep[see][for details and references]{1980lssu.book.....P, 
2005Natur.435..629S, 2009LNP...665..291V, 2011A&A...534A.128E}.

The physics of 
baryon acoustic oscillations 
in the hot plasma is also well
understood -- these oscillations were created by the interplay between the
gravitational pull of dark matter and the pressure of the photon-electron plasma in
the early universe \citep{1970ApJ...162..815P}.
BAO measurements together with cosmic microwave background (CMB) 
observations provide important tools to constrain
cosmological parameters and break degeneracies between them. 
Planck results \citep{2015arXiv150201589P} of the CMB
anisotropies have determined the scale of acoustic oscillations  
($109$~\Mpc) with high accuracy, in agreement with
several experimental BAO studies: 
SDSS \citep{2005ApJ...633..560E},
2dFGRS \citep{2005MNRAS.362..505C}, 
WiggleZ \citep{2011MNRAS.418.1707B}, 
6dF galaxy survey \citep{2011MNRAS.416.3017B}, SDSS-DR7 \citep{2012MNRAS.427.2132P}, 
and BOSS-DR9 \citep{2012MNRAS.427.3435A}. 
An overview of observable signatures of BAO in the galaxy distribution 
in the cosmic web can be found in \citet[][and references therein]{2013PhR...530...87W}.

Baryonic acoustic oscillations create weak shells in the distribution of galaxies.
The signatures of BAO have typically been detected with a two-point correlation function
\citep{2005ApJ...633..560E, 2009ApJ...696L..93M, 2013PhR...530...87W},
but other methods have also been proposed. \citet{2013MNRAS.429.1206J}
showed that the skewness of the cosmic density field contains an  
imprint of baryonic acoustic oscillations.
Recently,
\citet{2012A&A...542A..34A} proposed a new method for searching for BAO
shells in the distribution of galaxies -- a baolet analysis, which is a dedicated wavelet technique. BAO shells are shell-like systems of galaxies whose
central density enhancement is surrounded by an underdense region and then
higher density shell-like structures of galaxies. 
\citet{2012A&A...542A..34A}  searched for BAO shells using LRGs
(luminous red galaxies) as a centre candidate sample and the SDSS MAIN galaxy
sample to trace the shells. They detected BAO shells in the distribution
of galaxies.

Early studies of the cosmic web 
have shown the presence of giant voids of a scale of about $100$~\Mpc\  in our cosmic neighbourhood
\citep{1978IAUS...79..241J, 1982Natur.300..407Z}.
\citet{1994MNRAS.269..301E}  showed the presence of 
an $\approx 120 - 130$~\Mpc\ 
characteristic scale in the three-dimensional 
distribution of rich clusters and superclusters
of galaxies. Over- and underdensity regions of $200$~\Mpc\ and even $400$~\Mpc\
size have been detected in the quasar distribution \citep{2004MNRAS.355..385M,
2014A&A...568A..46E}.

Recently, \citet[][hereafter E2015]{2015A&A...580A..69E}
detected two peaks in the distance 
distribution of rich galaxy groups and clusters 
at $\approx 120$~\Mpc\ and at $\approx 240$~\Mpc\ around the
Abell cluster A2142. 

These results give rise to the question about the geometrical properties 
of the cosmic web, and, in particular, the characteristic scales 
of the structures in the cosmic web.  
In this paper we attempt to
answer this question by searching for the presence of characteristic
scales in the distance distribution between galaxy groups and
clusters that are expressed by the presence of maxima in these distributions, which
may be caused by the shell-like
structures around rich clusters in our cosmic neighbourhood. 
We use the sample of nearby rich clusters as 
shell centre candidates and analyse the space distribution of 
groups and clusters of galaxies in their neighbourhood
with the aim to find structures that resemble shells.
The detection of these structures  enables us to study
the properties of these shells and objects by tracing them in detail.

In Sect.~\ref{sect:data} we describe the 
data we used, in Sect.~\ref{sect:results} we give the results. We discuss the 
results and draw conclusions in Sect.~\ref{sect:discussion}.
At \url{http://www.aai.ee/~maret/Shell.html}
we present an interactive 3D model showing the distribution of
galaxy groups in the volume under study.

We assume  the standard cosmological parameters: the Hubble parameter $H_0=100~ 
h$ km~s$^{-1}$ Mpc$^{-1}$, the matter density $\Omega_{\rm m} = 0.27$, and the 
dark energy density $\Omega_{\Lambda} = 0.73$.

\section{Data} 
\label{sect:data} 

Our galaxy sample is the MAIN sample of the 8th data release 
of the Sloan Digital Sky 
Survey \citep{2011ApJS..193...29A} with the apparent $r$ magnitudes $r \leq 
17.77$, and the redshifts $0.009 \leq z \leq 0.200$, in total 576493 galaxies. 
We corrected the redshifts of galaxies for the motion relative to the CMB and 
computed the co-moving distances \citep{mar03} of galaxies. The absolute 
magnitudes of galaxies were determined in the $r$-band ($M_r$) with the $k$-
corrections for the SDSS galaxies, calculated using the KCORRECT algorithm 
\citep{2007AJ....133..734B}. We also applied evolution corrections, 
using the luminosity evolution model of \citet{blanton03b}. The magnitudes 
correspond to the rest-frame at the redshift $z=0$. 

Groups of galaxies were determined using the friends-of-friends cluster analysis 
method introduced in cosmology by \citet{tg76,1982Natur.300..407Z}. A galaxy belongs to a 
group of galaxies if this galaxy has at least one group member galaxy closer 
than a linking length. 
The aim in \citet{2012A&A...540A.106T, 2014A&A...566A...1T} was to find as
many groups as possible while keeping their properties uniform with respect to 
distance. 
In a flux-limited sample the density of galaxies slowly 
decreases with distance. To take this selection effect properly 
into account when constructing a group catalogue from a flux-limited sample, the 
linking length was rescaled  with distance, calibrating the scaling relation by observed 
groups. 
Therefore, the values of the linking length in physical coordinates 
in the group catalogue depend on the distance. In the redshift interval
$0.04 < z < 0.12$, used in our study (see below), 
these values are approximately $0.3 - 0.5$~\Mpc.
As a result, the 
largest sizes in the sky projection and the velocity dispersions of our groups 
are similar at all distances. Since our study is focused on rich groups
of galaxies, it is important to note that with these values of the 
linking length, our groups do not include large sections of surrounding
galaxy filaments or superclusters 
\citep[see][for detailed discussion]{2014A&A...566A...1T}.
The details of the data 
reduction and the description of the group catalogue can be found in 
\citet{2012A&A...540A.106T, 2014A&A...566A...1T}.
When compiling the group catalogue, various selection effects that might 
affect the catalogue were taken into account.
In particular, the SDSS galaxy sample is incomplete because of fibre 
collisions - the smallest separation between spectroscopic fibres is 55",
and about 6\% of galaxies in the SDSS are without observed spectra. 
\citet{2012A&A...540A.106T, 2014A&A...566A...1T} studied the effect of missing galaxies
on a group catalogue and concluded that this mostly affects galaxy pairs.
They estimated that approximately 8\% of galaxy pairs may be missing from the catalogue.
This does not affect our study, which is mainly focused on richer galaxy groups.
We used the data of groups and clusters in the comoving
distance interval 120~\Mpc\ $\le D \le $ 340~\Mpc\ (corresponding to 
the redshift range $0.04 < z < 0.12$) 
where the selection effects are the smallest \citep[we discuss the 
selection effects in detail in ][]{ 2012A&A...542A..36E}.
As candidates of  shell centres we chose data of groups with at least 50 member 
galaxies that were analysed previously in \citet{2012A&A...540A.123E, 2012A&A...542A..36E}.
They correspond to rich galaxy clusters.  This sample 
of 109  clusters includes all clusters from the SDSS DR8 with at least 50 member 
galaxies in this distance interval. 

We calculated the distance distributions and radial densities
from these rich clusters to other 
groups and clusters of different richness from richest clusters to poorest
groups (galaxy pairs) in our catalogue and searched for a maxima in these
distributions. These indicate the possible presence of shell-like
structures in the galaxy distribution. 
To estimate the effect of the errors in distance distributions, 
we applied the following procedure. 
We determined the distance distributions by the kernel method, using the
B3 box spline kernel 
\citep[see e.g. ][]{2012A&A...539A..80L}
and selecting the kernel width as recommended by \citet{1986desd.book.....S}.  
The kernel width depends on both the variance of the distance distribution 
and the number of groups, therefore we used different kernel widths for different richness classes, 
in the range of $7.5$~\Mpc\  to $28.3$~\Mpc.
To determine the density distributions, we generated reference random point distributions
(of 200000 points) for every richness class, 
following the SDSS mask and the radial selection function for groups of this class. 
The density distribution is the ratio of the observed and the 
reference distance distributions. We found bootstrap errors of distance and density
distributions, using the smoothed bootstrap method
\citep[see, for example, ][]{davhink97}.
This is based on the idea of smoothing the observed integral probability 
distribution for bootstrap selection 
and is realised by selecting the observed points with replacement, 
as usual, and adding to their coordinates a random replacement 
with the probability distribution given by the kernel function used, 
but with a smaller width (we used half of the original width).  
We also applied a Kolmogorov-Smirnov test to compare different
distance distributions.
Errors of the distances between clusters are small, of the order of the cluster
sizes (approximately $1$~\Mpc), and their influence on the results of 
the statistical tests is negligible.
With the Kolmogorov-Smirnov test we compared the integral distribution
of the data without binning, 
applying an empirical cumulative distribution function to the data.

\section{Results}
\label{sect:results} 

\subsection{Distances between rich clusters and groups}
\label{sect:alldist}

\begin{figure}[ht]
\centering
\resizebox{0.40\textwidth}{!}{\includegraphics[angle=0]{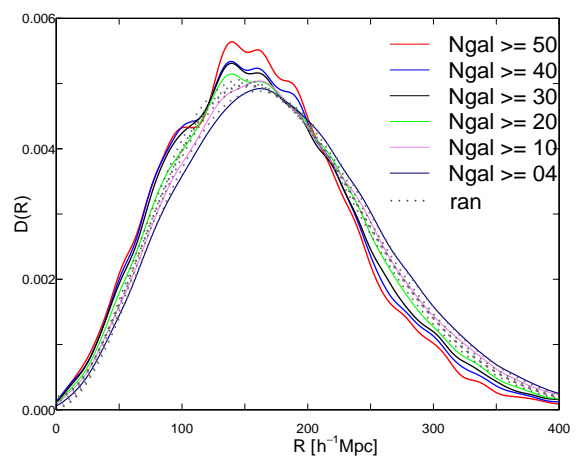}}
\caption{
Distance distributions from all rich clusters
with at least 50 member galaxies to other
groups and clusters of different richness and to random points
(see figure legend and text for details; random catalogues
were generated for every richness class, but since the distributions
are overlapping, we plot them all with dotted grey lines).
}
\label{fig:alldistd}
\end{figure}

We calculated distances from rich clusters with at least
50 member galaxies to clusters and groups
with $N_{\mathrm{gal}} \geq 50$ (109), 
$N_{\mathrm{gal}} \geq 40$ (176),
$N_{\mathrm{gal}} \geq 30$ (310),
$N_{\mathrm{gal}} \geq 20$ (583),
$N_{\mathrm{gal}} \geq 10$ (1972), 
$N_{\mathrm{gal}} \geq 4$ (11084), and
$N_{\mathrm{gal}} \geq 2$ (45858) (we give the number of systems
in parenthesis).
Figure~\ref{fig:alldistd} shows the distribution of the distances from
all rich clusters with at least 50 member galaxies 
together with distances from rich clusters to random points. 
All curves in Fig.~\ref{fig:alldistd} are normalised so that each integrates to $1$.
We looked for specific features in the distance   
distributions between groups.  
The main form of the distribution is caused by the shallowness of our sample
(its thickness is only $220$~\Mpc,\ while it is $\approx 600$~\Mpc\ wide).
Several wiggles seen in the distance distributions 
indicate possible density maxima around the clusters. 
One with the highest height is located at distance of
$130$~\Mpc,\ indicating an enhanced density of groups and clusters around
rich clusters at this distance. The wiggles are seen in distributions
of distances of rich clusters and clusters with 20 -- 50 member
galaxies and disappear from distributions
if calculations include groups with 2 -- 10 member galaxies. This 
suggests that possible structures with enhanced density are 
more clearly delineated by rich groups and clusters of galaxies.  
The Kolmogorov-Smirnov test, applied to the distance distributions  
in the distance interval of 
$90 - 140$~\Mpc,\
confirmed that the differences in distributions of distances from rich clusters
with at least 50 member galaxies to groups with 2 -- 10 member
galaxies are statistically highly significant, with a $p$-value $p < 0.01$.
Wiggles like this are not seen in the distance distributions to random points.
As seen in Fig.~\ref{fig:alldistd}, the distance distributions
of random points and poor groups overlap, showing that  
these distributions are similar.

\begin{figure}[ht]
\centering
\resizebox{0.45\textwidth}{!}{\includegraphics[angle=0]{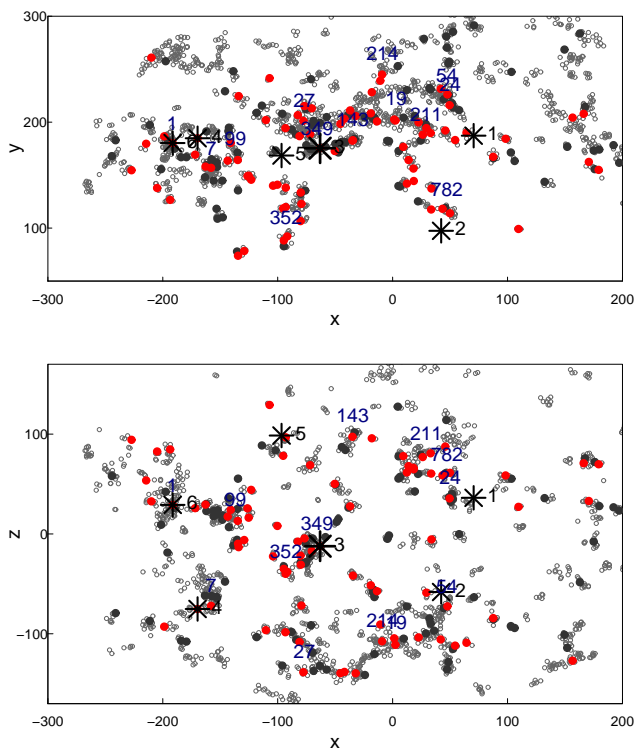}}
\caption{
Distribution  of supercluster member groups with at least
four galaxies in Cartesian coordinates 
as defined in the text, in $h^{-1}$ Mpc. 
Grey empty circles denote groups with at least four member galaxies,  
grey filled circles denote clusters with $30 \leq N_{\mathrm{gal}} \leq 50$, and
red filled circles mark the location of clusters with $N_{\mathrm{gal}} \geq 50$.
Black stars denote clusters that might be centres of enhanced density structures, as
explained in the text, black numbers are order numbers of possible centre clusters 
from Table~\ref{tab:cl7data}. Blue numbers are supercluster ID numbers from
Table~\ref{tab:scl}.
}
\label{fig:cl5030xyz}
\end{figure}

In Fig.~\ref{fig:cl5030xyz} we plot the distribution of groups and clusters
with at least four member galaxies in  Cartesian
coordinates 
defined as in \citet{2007ApJ...658..898P} and in \citet{2012A&A...539A..80L}:
$x = -d \sin\lambda$, $y = d \cos\lambda \cos \eta$, and
$z = d \cos\lambda \sin \eta$, 
where $d$ is the comoving distance, and $\lambda$ and $\eta$ are the SDSS 
survey coordinates. In the SDSS, the survey coordinates form
a spherical coordinate system, 
where $(\eta,\lambda) = (0,90.)$ corresponds to $(R.A.,Dec.) = (275.,0.)$, 
$(\eta,\lambda) = (57.5,0.)$ corresponds to $(R.A.,Dec.) = (0.,90.)$,
and at  $(\eta,\lambda) = (0.,0.)$,  $(R.A.,Dec.) = (185.,32.5)$. 
Here the locations of all rich groups and clusters
with at least 30 member galaxies are plotted, but to avoid strong 
projection effects, we show only those
poor groups that are supercluster members. 
Rich clusters of galaxies with at least 50 members are plotted in red.
This figure shows that many rich clusters 
are located at the $y$-coordinate interval approximately  $170 - 220$~\Mpc,
at which rich galaxy superclusters, including those belonging to the
Sloan Great Wall, form three chains of superclusters,
separated by voids of size of about $100$~\Mpc.
The  Bootes supercluster (Scl~349) with its richest cluster, 
A1795, is located in the middle chain of superclusters. 
We also see rich clusters in the chains of nearby superclusters  
that connect the Hercules supercluster and the chains of rich superclusters
behind it \citep[the distribution of galaxy superclusters was described in 
detail in][]{2011A&A...532A...5E}.

As a next step, 
we analysed the distance distributions of groups and clusters
of galaxies around each rich cluster individually.
This analysis showed that whether the distance distribution has a maximum or not
depends on the location of the cluster in the cosmic web (Fig.~\ref{fig:cl5030xyz}).
Cluster A1795 in the middle chain of superclusters in the Bootes
supercluster (SCl~349, cluster 3 in the figure
and in Table~\ref{tab:cl7data}) 
has the strongest maximum in the distance distributions.
In the next section we describe in detail the structures
in the cosmic web that give rise to these maxima.
We found 28 galaxy clusters with 
a  maximum at $120 - 130$~\Mpc\ 
in these distributions.
They can be considered as shell centre candidates. 
If these clusters are located in the same (or neighbouring) 
superclusters, they are grouped together and represent the same possible
centre, therefore we selected one cluster per possible centre.
Data on six shell centre candidate clusters are given in Table~\ref{tab:cl7data}.
We list the coordinates, distances, total luminosities,
and the peak values of the luminosity density field at the location of the cluster
\citep[at a smoothing length  of $8$~\Mpc, calculated with the $B_3$ spline kernel
in units of mean luminosity density, $\ell_{\mathrm{mean}}$ = 
1.65$\cdot10^{-2}$ $\frac{10^{10} h^{-2} L_\odot}{(\vmh)^3}$][]{2012A&A...539A..80L}.
Next we describe the maximum, and the spatial structure around cluster A1795 in detail. 
Data on galaxy superclusters mentioned in the text can be found 
in Table~\ref{tab:scl} \citep[see][for details]{2012A&A...539A..80L, 2012A&A...542A..36E}.

\begin{table*}[ht]
\caption{Data on  shell centre candidate clusters.}
\begin{tabular}{rrrrrrrrrr} 
\hline\hline  
(1)&(2)&(3)&(4)&(5)& (6)&(7)&(8)&(9)&(10) \\      
\hline 
 No.&ID&$N_{\mathrm{gal}}$& $\mathrm{R.A.}$ & $\mathrm{Dec.}$ &
 $\mathrm{Dist.}$ & $L_{\mathrm{tot}}$ & $D8$ & Abell ID & SCl ID\\
 && &[deg]&[deg]&[$h^{-1}$ Mpc]&[$10^{10} h^{-2} L_{\sun}$]& & & \\
\hline
1& 20159 &  52 & 158.02 & 40.16 & 203.31 &  59.80  & 3.9  & \object{A 1026}, \object{A 1035}X   &  ---   \\
2&  3714 &  82 & 164.58 &  1.56 & 121.07 &  54.71  & 5.6  & \object{A 1139}X                    & ---  \\
3& 23374 & 114 & 207.22 & 26.68 & 186.81 & 100.95  & 9.0  & \object{A 1795}X, \object{A 1818}& 349\\
4& 34513 &  53 & 225.86 &  7.88 & 261.94 &  94.18  & 5.9  & \object{A 2020}               & 1311\\
5& 50647 &  52 & 232.32 & 52.88 & 217.84 &  62.30  & 4.4  &   --- & --- \\                       
6& 29587 & 207 & 239.52 & 27.32 & 264.53 & 365.47  & 20.7 & \object{A 2142}X                    & 1  \\
\label{tab:cl7data}                                                          
\end{tabular}\\
\tablefoot{                                                                                 
Columns are as follows:
1: Order number  of the cluster;
2: ID of the cluster;
3: the number of galaxies in the cluster, $N_{\mathrm{gal}}$;
4--5: cluster right ascension and declination, ($\mathrm{R.A.}$ and $\mathrm{Dec.}$);
6: cluster comoving distance, $\mathrm{Dist.}$;
7: cluster total luminosity, $L_{\mathrm{tot}}$;
8: the peak value of the luminosity density field at the location of the cluster, $D8$;
9: Abell ID of the cluster; X denotes X-ray cluster \citep{2003AJ....126.2740L};
10: SCl ID of the cluster from \citet{2012A&A...539A..80L}.
}
\end{table*}

\begin{figure}[ht]
\centering
\resizebox{0.45\textwidth}{!}{\includegraphics[angle=0]{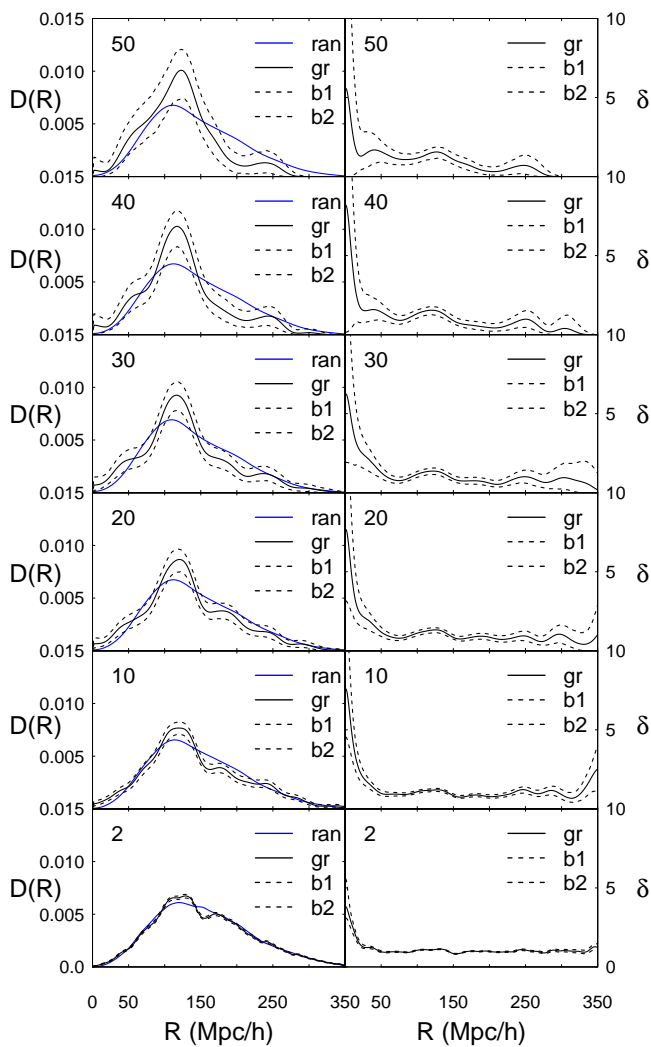}}
\caption{
Left panels show the distribution of distances between cluster A1795 and other
groups and clusters of different richness
from pairs to clusters with $N_{\mathrm{gal}} \geq 50$. Right panels
show the radial density profile around A1795
($\delta(r) = \frac{\rho(r)}{\rho_0}$, 
where $\rho_0$ is the average density of the sample). Richness limits of groups
and clusters are given in the panels. The legends are: gr stands for group
distributions, ran for the random samples, and b1 and b2 denote the upper and lower 95\% 
bootstrap confidence limits for the group distributions.
}
\label{fig:gr23374distd}
\end{figure}

\subsection{Structure around cluster A1795}
\label{sect:a1795} 

We show in
Fig.~\ref{fig:gr23374distd} the distributions of distances of rich and
poor clusters and groups from 
cluster A1795. We also plot the distance and density distributions of the random samples
that follow the mean dependence of the groups on distance from the observer
(the reference distribution) and the 95\% 
confidence limits for the distributions, obtained by smoothed bootstrap, 
as described above.
Random  
distributions are smooth with a maximum around $150$~\Mpc\ that is  
caused by the geometry of the sample (Fig.~\ref{fig:gr23374distd}).
The selection functions used to generate the random distributions account 
for the fact that the spatial density of groups of various richness 
in our sample differs slightly. Thus the dependence  of the density of groups 
of different richness in the random distributions on 
distance  follows that in our group catalogue.

\begin{table}[ht]
\caption{Data of superclusters.}
\begin{tabular}{rrrrrr} 
\hline\hline  
(1)&(2)&(3)&(4)&(5)& (6)\\      
\hline 
 ID(long) &ID &   $Dist.$&$L_{\mathrm{tot}}$ & $D_{\mathrm{peak}}$ & ID(E01)\\
\hline
239+027+0091 &   1 &  264 & 1809   &  22.2   & ---    \\
227+006+0078 &   7 &  233 & 1675   &  17.0   & 154    \\
184+003+0077 &  19 &  231 & 2919   &  15.0   & 111    \\
167+040+0078 &  24 &  225 &  751   &  14.6   & 95     \\
202-001+0084 &  27 &  256 & 5163   &  14.0   & 126    \\
173+014+0082 &  54 &  241 & 2064   &  12.4   & 111    \\
230+027+0070 &  99 &  215 & 2874   &  11.5   & 158    \\
229+006+0102 & 103 &  302 & 1004   &  11.0   & 160    \\
135+038+0094 & 124 &  281 &  548   &  10.5   & 249    \\
152-000+0096 & 126 &  285 & 1001   &  10.1   & 82     \\
203+059+0072 & 143 &  211 &  753   &  10.1   & 133    \\
172+054+0071 & 211 &  207 & 1618   &   9.2   & 109    \\
216+016+0051 & 349 &  159 &  284   &   8.5   & 138    \\
227+007+0045 & 352 &  135 &  379   &   8.7   & 154    \\
151+054+0047 & 782 &  139 &  465   &   6.8   & ---    \\
226+007+0090 &1311 &  272 &   95   &   5.4   & ---    \\

\label{tab:scl}  
\end{tabular}\\
\tablefoot{                                                                                 
The columns are as follows:
1: Long ID of a supercluster AAA+BBB+ZZZZ, 
where AAA is R.A., +/-BBB is Dec. (in degrees), and ZZZZ is 1000$z$ 
\citep[see][for details]{2012A&A...539A..80L};
2: ID of a supercluster from the SDSS DR8 superclusters catalogue;
3: the distance of a supercluster, in $h^{-1}$~Mpc;
4: the total weighted luminosity of galaxies in the supercluster, $L_{\mbox{tot}}$,
in $10^{10} h^{-2} L_{\sun}$;
5: the luminosity density value at the density maximum (density contrast), $d_{\mbox{peak}}$, 
in units of the mean luminosity density \citep{2012A&A...539A..80L};
6: ID(E01): the supercluster ID  in  the catalogue by \citet{2001AJ....122.2222E}. SCl~160 -- the Hercules
supercluster, SCl~111 and SCl~126 -- superclusters of the Sloan Great Wall, SCl~158 -- the Corona 
Borealis supercluster, SCl~138 -- the Bootes supercluster, SCl~109 -- the Ursa 
Major supercluster.
}
\end{table}

Figure~\ref{fig:gr23374distd} shows in the distributions of distances
between A1795 and other rich clusters with 
$N_{\mathrm{gal}} \geq 30$ at least two maxima, 
the first  maximum at the distance  approximately  $120$~\Mpc,
and a second one, at the distance  approximately  $240$~\Mpc\
($122$ and $238$~\Mpc\ exactly).
Out of 108 rich clusters in our sample, 54 are located
in the distance interval of $90 \leq D \leq 140$~\Mpc\ from A1795.
The bootstrap analysis and the comparison with random
distribution (blue line in Fig.~\ref{fig:gr23374distd}, and the 
Kolmogorov-Smirnov test for the full samples) 
confirmed that the observed distance distributions are different 
from random distributions at a high significance level.
The location of the maximum does not depend on
richness, but its height increases with the group richness. 
The density distributions show minimum and 
maximum at the same scales. 
There is a slight increase of densities and bump in the distance distributions
at scales smaller than $47$~\Mpc.
At the density maximum at $120$~\Mpc,\ the cluster density
is about 1.6 times higher than the mean cluster density in the sample.  

The tests mentioned above showed that observed distributions
differ from random distributions, 
but as the number of rich groups in our sample is small, 
 the maximum we find might still be random fluctuations. 
They also show a difference from random 
samples, but this does not say whether the difference is related to the presence of maxima
in the distance distributions. 
To check this, we generated 1000 random samples for every richness class 
with the same number of groups as in the corresponding class. 
These samples populate the same volume as the observed clusters and have 
the same selection function. For each of these random samples, 
we found the distance and density distribution around A1795 in
the same way 
as we did for the observed sample and determined 
the maxima in these distributions.
The mean number of maxima depended on the richness class, 
being about 4 for the groups of richness $N_{\mathrm{gal}} \geq 50$ 
and about 6 for $N_{\mathrm{gal}} \geq 20$. 
We found the probability distribution of these maxima 
in the $(R_{max}, D_{max})$ plane and compared it with 
the values of $(R_{max}, D_{max})$ for the observed maxima. 
This gave us the probability of obtaining maxima in the random distribution.
By comparing the distribution of these maxima, 
we found a few random maxima near the observed maxima 
at small ($~40$~\Mpc) and large ($~240$~\Mpc) distances, 
but the maximum around $120$~\Mpc\ lay far outside 
the minimum (1/1000) probability contour for the random maxima. 
We can therefore say that this maximum is real, 
with the probability that we obtain such a maximum from random
distributions lower than 0.001, 
but the other maxima might be caused by the small number of groups in the sample.

Next we analyse the space distribution of 
galaxy groups and clusters around cluster A1795 to understand 
which structures are behind the minima and maxima in the distance and density
distributions. 

\begin{figure}[ht]
\centering
\resizebox{0.45\textwidth}{!}{\includegraphics[angle=0]{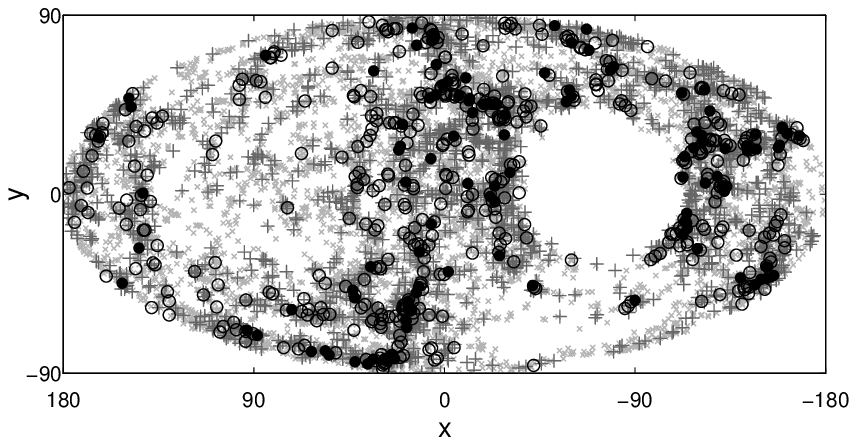}}
\resizebox{0.45\textwidth}{!}{\includegraphics[angle=0]{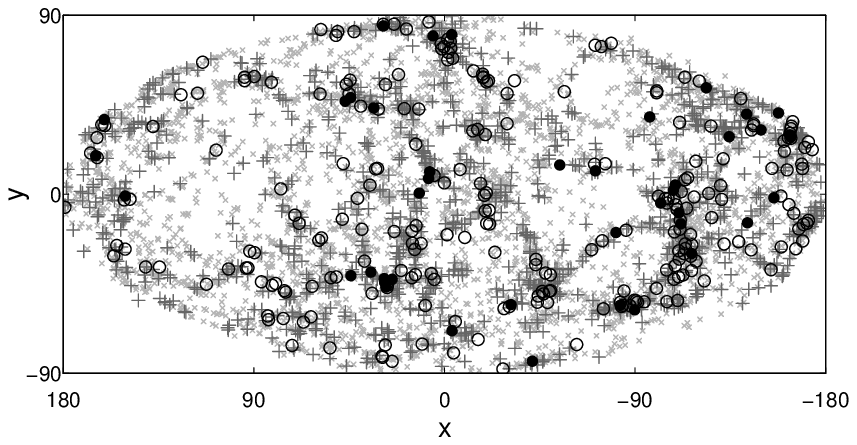}}
\caption{
Sky distribution of galaxy groups around cluster A1795.
$X$ and $Y$ are sky coordinates in a Hammer-Aitoff projection
\citep{mar03}. 
Black filled circles denote clusters with $N_{\mathrm{gal}} \geq 30$,
black empty circles those with $10 \geq N_{\mathrm{gal}} < 30$, 
grey crosses those with  $4 \geq N_{\mathrm{gal}} < 10$, and grey dots
show poor groups with only two or three member galaxies.
Upper panel: groups and clusters in a distance interval of 
$90 - 130$~\Mpc, lower panel: in a distance interval of 
$50 - 90$~\Mpc. The void in the upper panel is due to
the boundary effects.
}
\label{fig:gr23374sky}
\end{figure}

Figure~\ref{fig:gr23374sky} shows the sky distribution of poor and rich groups
around cluster A1795 in a distance interval of 90~\Mpc\ $\le D \le $ 130~\Mpc\
(upper panel), and in a distance interval of 50~\Mpc\ $\le D \le $ 90~\Mpc\
(lower panel).
Almost the complete sky is covered; only a small part of the shell is not 
visible as a result of the sample boundaries. 
This plot shows filaments of various richness  in the shell. 
In the upper panel one of the richest  structures is the Sloan 
Great Wall (the superclusters SCl~027 and SCl~019)
in the lower central part of the plot, delineated by rich
galaxy clusters.  
Here, in the region of the Sloan Great Wall, the highest luminosity density contrast
value is $D8 = 15.0$, see Table~\ref{tab:scl}. 
In contrast, in the middle of the left part of the
figure we see a low-density region that is only crossed by filaments of poor
groups. This clearly shows that the shell is determined by filamentary galaxy 
systems of various richness. 
The lower panel also shows filamentary galaxy systems, but they are
generally poor. Even the richest systems here consist mostly of 
poor groups and clusters of intermediate 
richness.

\begin{figure}[ht]
\centering
\resizebox{0.45\textwidth}{!}{\includegraphics[angle=0]{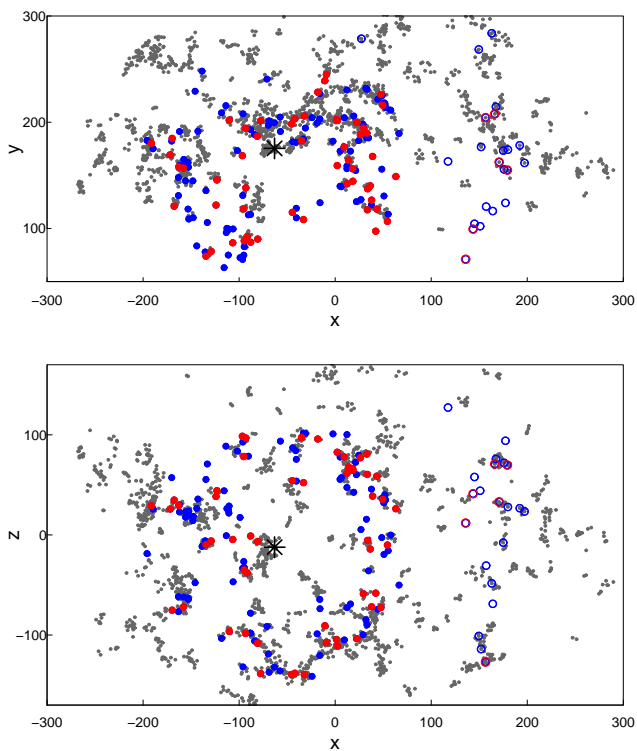}}
\caption{
Distribution  of supercluster member groups with at least
four galaxies in Cartesian coordinates (in $h^{-1}$ Mpc,
grey dots). 
Red filled circles denote clusters with $N_{\mathrm{gal}} \geq 50$, and
blue filled circles denote clusters with $30 \leq N_{\mathrm{gal}} \leq 50$
in a distance interval of $90 \leq D \leq 140$~\Mpc\ from A1795.
Red empty circles denote clusters with $N_{\mathrm{gal}} \geq 50$, and
blue empty circles denote clusters with $30 \leq N_{\mathrm{gal}} \leq 50$
in a distance interval of $200 \leq D \leq 240$~\Mpc\ from A1795.
The star denote the location of A1795. 
}
\label{fig:gr23374xyz}
\end{figure}

To understand the space distribution of galaxy groups and clusters 
in this region better, we show in 
Fig.~\ref{fig:gr23374xyz} the distribution of groups and clusters
with at least four member galaxies in  Cartesian
coordinates. 
Below we describe the structures of the cosmic web around A1795 in detail.
The central part of the structure seen in the figure
is formed by cluster A1795 and other groups and clusters in the Bootes
supercluster (SCl~349 with a highest luminosity
density $D = 8.5$). 
They are surrounded by underdense regions. 
Chains of galaxy systems penetrating these voids
connect the Bootes supercluster with other superclusters that
form 
the walls of the shell. These chains cause the density enhancement
at distances $D < 70$~\Mpc\ seen in Fig.~\ref{fig:gr23374distd}
and in the sky distribution in Fig.~\ref{fig:gr23374sky}. After a minimum at 
about $D \approx 80$~\Mpc,\ the
next maximum in the distance distribution at $D \approx 120$~\Mpc\
is due to clusters and groups in rich superclusters. 
ID numbers of some superclusters are shown 
 in Fig.~\ref{fig:gr23374xyz}.
They include the superclusters
from the Sloan Great Wall, SCl~027 and SCl~019 with a density contrast $D8 = 14.04$ and $15.0$
in the lower part of
the lower panel, the Ursa Major supercluster (SCl~211) with $D8 = 9.2$
and SCl~024 with $D8 = 14.6$ near SCl~211 in the upper part of the right panel,
and the superclusters SCl~099 (the Corona Borealis supercluster with $D8 = 11.5$)
and SCl~001 with $D8 = 22.2$ on the left side of panels. 
The superclusters SCl~352 with $D8 = 8.7$ and SCl~782 with $D8 = 6.8$
are members of galaxy filaments
that penetrate the void between the nearby rich Hercules supercluster ($D8 = 11.0$)
and the superclusters SCl~027 and SCl~211, marking the
other edges of the shell. 
Thus, galaxy superclusters around cluster A1795
form a structure close to a spherical shell-like structure since 
rich superclusters, connected by
galaxy filaments  and poor superclusters, are approximately
at the same distance from the centre cluster A1795.
The maximum in the distance distribution is wide, 
which is due to the width of superclusters.
The  voids are not empty but marked by
poor galaxy groups (Fig.~\ref{fig:gr23374sky}). 
This shows that the overall distribution of the poor groups
is smoother than that of rich clusters, and their density contrast
is lower, as seen in Fig.~\ref{fig:gr23374distd}.
The distribution of rich galaxy superclusters
in this region was also described in \citet{2011A&A...532A...5E}.

Groups and clusters at distances $D > 220$~\Mpc\
are located across the void region at a distance of $120$~\Mpc\
from the wall  surrounding cluster A1795. Possibly,
they form another shell. However, the far
part of this possible wall are not covered by SDSS, and we cannot study this 
structure in detail.

\begin{figure}[ht]
\centering
\resizebox{0.22\textwidth}{!}{\includegraphics[angle=0]{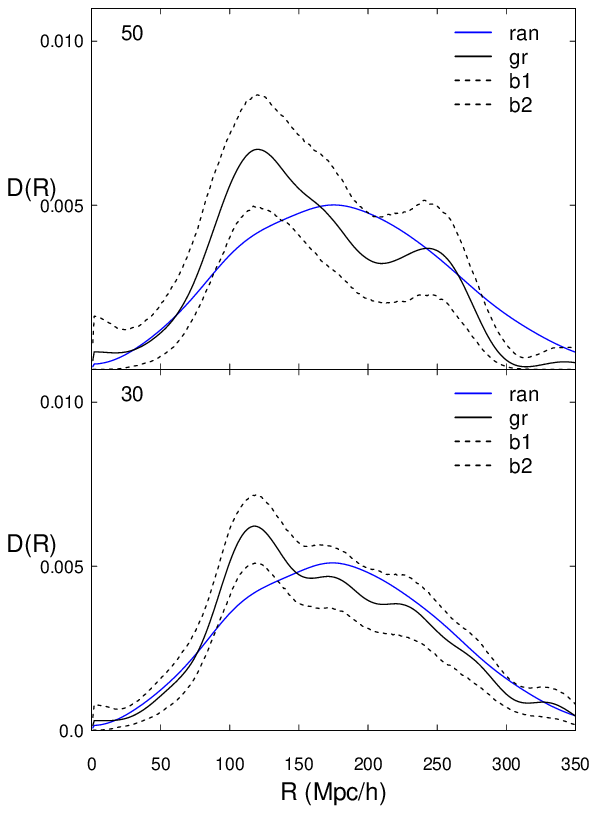}}
\resizebox{0.25\textwidth}{!}{\includegraphics[angle=0]{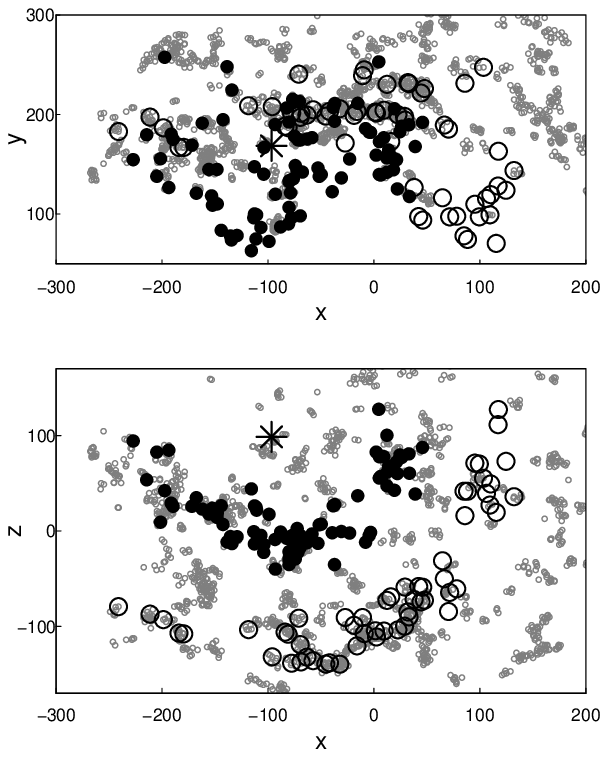}}
\caption{
Distance distribution between cluster Gr50647 and other
groups and clusters of different richness (left panels)
and the distribution of groups with at least
four galaxies in Cartesian  coordinates
(right panels). 
The richness limits of groups
and clusters are given in the left panels, gr stands for group
distributions, ran for the random samples, and b1 and b2 denote the upper and lower 95\% 
bootstrap confidence limits for the group distributions.
In the right panels grey empty circles show groups with at least four member galaxies.  
Black filled circles denote clusters with $N_{\mathrm{gal}} \geq 30$
in a distance interval 
of 90~\Mpc\ $\le D \le $ 140~\Mpc\ from Gr50647, and
black empty circles those with $N_{\mathrm{gal}} \geq 30$
in a distance interval 
of 200~\Mpc\ $\le D \le $ 250~\Mpc\ from Gr50647.
The star denotes the location of Gr50647. 
The positions of the maxima are $119$ and $242$~\Mpc\ (upper left panel, 
$N_{\mathrm{gal}} \geq 50$), and $117$ and $219$~\Mpc\ (lower left panel, 
$N_{\mathrm{gal}} \geq 30$). 
} 
\label{fig:gr50647}
\end{figure}

\subsection{Structures around other rich clusters}
\label{sect:multi} 

Our calculations showed five more possible shell candidates
for which central galaxy clusters
had a maximum at a scale of about $120 - 130$~\Mpc\  
in the distance distribution to other galaxy groups and
clusters. None of these possible shells are as fully embedded in our sample volume
as the shell around cluster A1795.
We applied the smoothed bootstrap test to compare the observed and random distance 
distributions and found for all these clusters that the observed 
distributions differ from random with high significance. With the maxima test
described above, we tested the probability that these maxima can also be found 
 in random distributions. 
Our sample is shallow,
the mean density run of the randomised samples does not represent the local
group density distribution well. We therefore did not use the normalised density as shown in 
Fig.~\ref{fig:gr23374distd}, but show the direct distance distributions instead.
The maximum in the distance distributions is not always at the same position.
Depending on the location of clusters of different richness
in superclusters and on the width of superclusters, 
the values of the maximum for these clusters are 
in a distance interval
of $117 - 136$~\Mpc\ in the distance distribution of clusters with 
$N_{\mathrm{gal}} \geq 30$. 
We list the values of distance maxima in the captions of the
distance distribution figures.
We also present the figures of the distribution of groups and clusters
with at least four member galaxies in  Cartesian
coordinates around these five clusters. 
The data of these clusters are listed in Table~\ref{tab:cl7data}.
Systems with
the number of member galaxies $N_{\mathrm{gal}} \geq 30$ are plotted
 with black filled circles if their
distances from the shell centre cluster lie 
in an interval of 100~\Mpc\ $\le D \le $ 150~\Mpc,
and with black empty circles if their distances lie in an
interval of between 200~\Mpc\ $\le D \le $ 250~\Mpc\ from the centre.

\begin{figure}[ht]
\centering
\resizebox{0.22\textwidth}{!}{\includegraphics[angle=0]{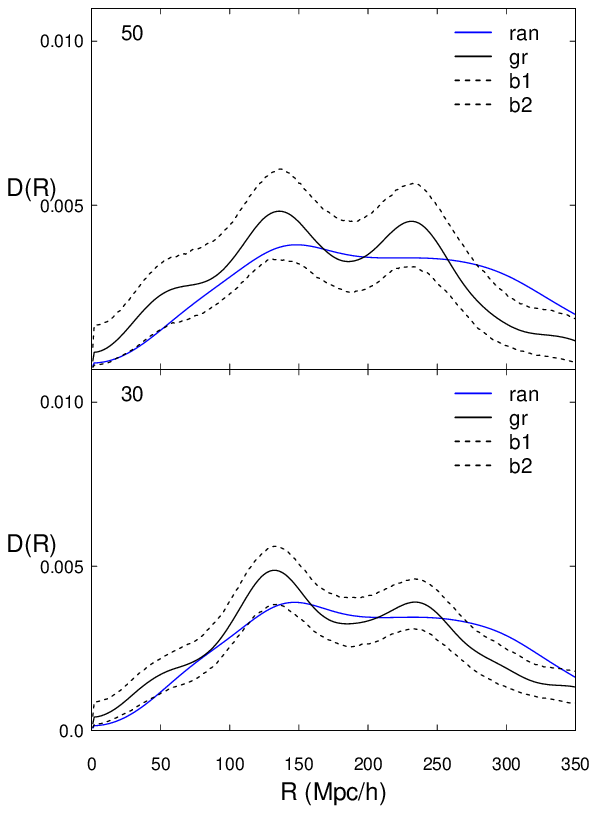}}
\resizebox{0.25\textwidth}{!}{\includegraphics[angle=0]{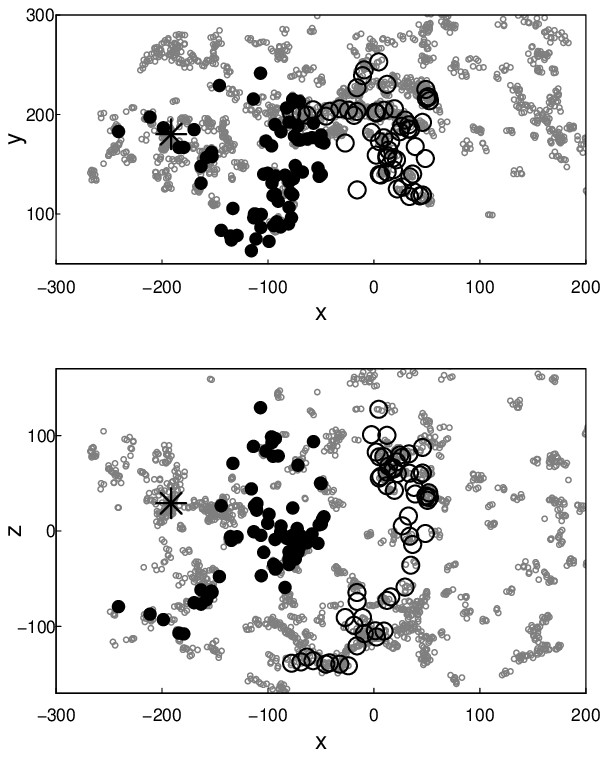}}
\caption{
Distance distribution between cluster A2142 (Gr29587) and other
groups and clusters of different richness (left panels),
and the distribution of groups with at least
four galaxies in Cartesian  coordinates
(right panels). Notations and details are the same as in Fig.~\ref{fig:gr50647}.
The maxima are located at $135$ and $231$~\Mpc\ (upper left panel, 
$N_{\mathrm{gal}} \geq 50$), and $131$ and $233$~\Mpc\ (lower left panel, 
$N_{\mathrm{gal}} \geq 30$). 
} 
\label{fig:gr29587}
\end{figure}

Cluster Gr50647 (Fig.~\ref{fig:gr50647}) 
lies at the edge of one chain of superclusters
described in \citet{2011A&A...532A...5E, 2012A&A...542A..36E},
the superclusters SCl~143 with $D8 = 10.1$
and SCl~211 (the Ursa Major supercluster with $D8 = 9.2$) are
among them.
The walls of the possible shell are formed by superclusters
in the nearby void region and another chain
of superclusters, including the Corona Borealis supercluster  with $D8 = 11.5$.
Clusters at distances $D > $ 200~\Mpc\ from Gr50647 lie in the Sloan Great Wall.
The maxima test confirmed that the maximum at $130$~\Mpc\
has a probability lower than 0.001 to be drawn from the
random distributions.

\begin{figure}[ht]
\centering
\resizebox{0.22\textwidth}{!}{\includegraphics[angle=0]{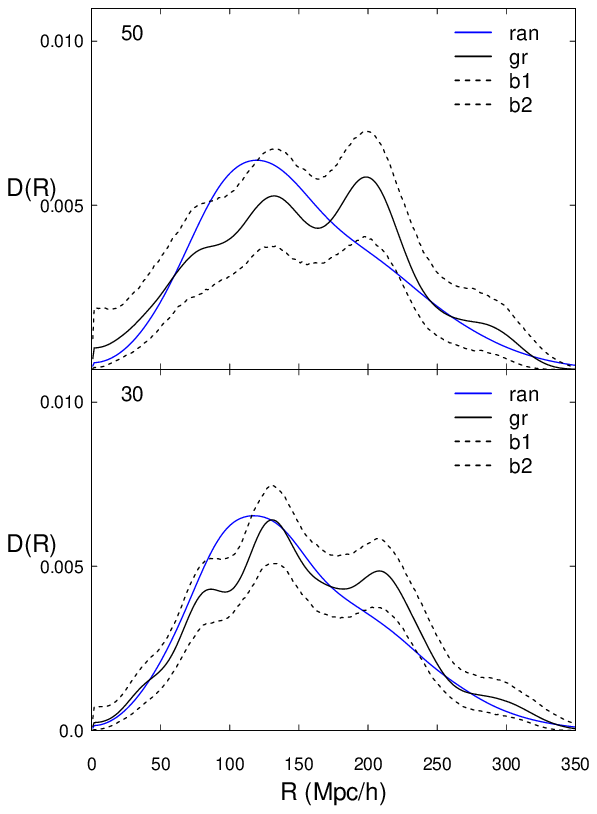}}
\resizebox{0.25\textwidth}{!}{\includegraphics[angle=0]{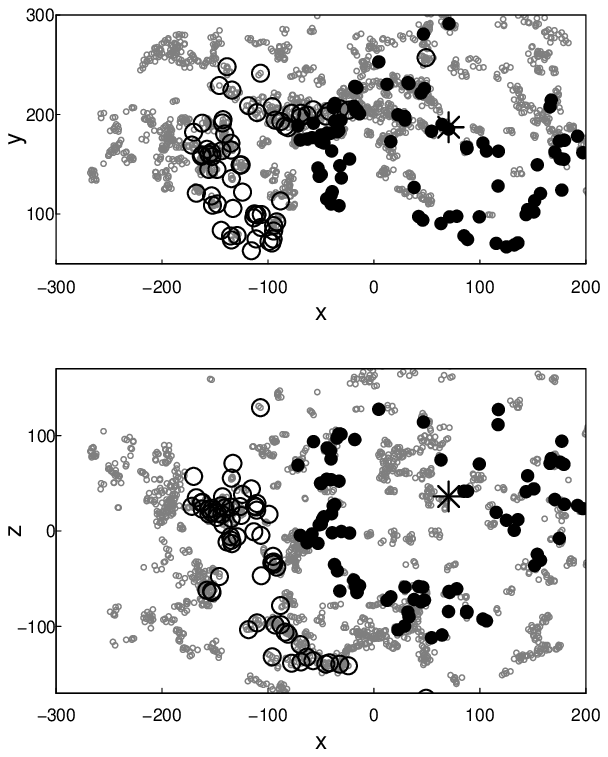}}
\caption{
Distance distribution between cluster Gr20159 (A1026) and other
groups and clusters of different richness (left panels)
and the distribution of groups with at least
four galaxies in Cartesian  coordinates
(right panels). Notations and details are the same as in Fig.~\ref{fig:gr50647}.
The maxima are located at $131$ (upper left panel, 
$N_{\mathrm{gal}} \geq 50$), and $130$ and $207$~\Mpc\ (lower left panel, 
$N_{\mathrm{gal}} \geq 30$). 
} 
\label{fig:gr20159}
\end{figure}

\begin{figure}[ht]
\centering
\resizebox{0.22\textwidth}{!}{\includegraphics[angle=0]{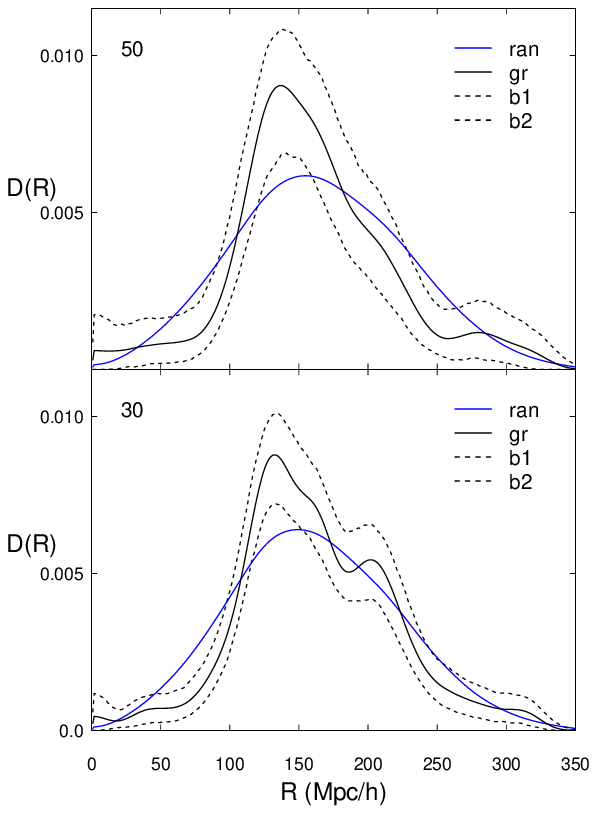}}
\resizebox{0.25\textwidth}{!}{\includegraphics[angle=0]{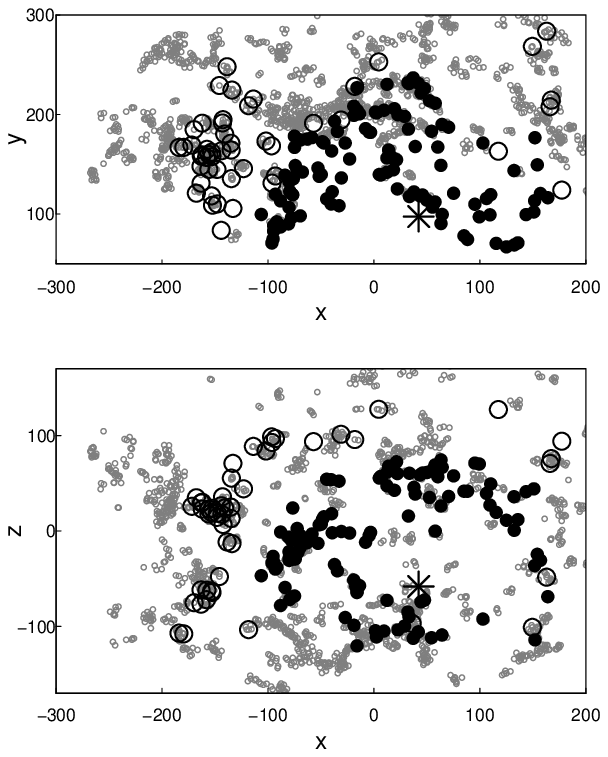}}
\caption{
Distance distribution between cluster Gr3714 (A1139) and other
groups and clusters of different richness (left panels)
and the distribution of groups with at least
four galaxies in Cartesian  coordinates
(right panels). Notations and details are the same as in Fig.~\ref{fig:gr50647}.
The maxima are located at $136$ and $244$~\Mpc\ (upper left panel, 
$N_{\mathrm{gal}} \geq 50$), and $131$ and $216$~\Mpc\ (lower left panel, 
$N_{\mathrm{gal}} \geq 30$). 
} 
\label{fig:gr3714}
\end{figure}

\begin{figure}[ht]
\centering
\resizebox{0.22\textwidth}{!}{\includegraphics[angle=0]{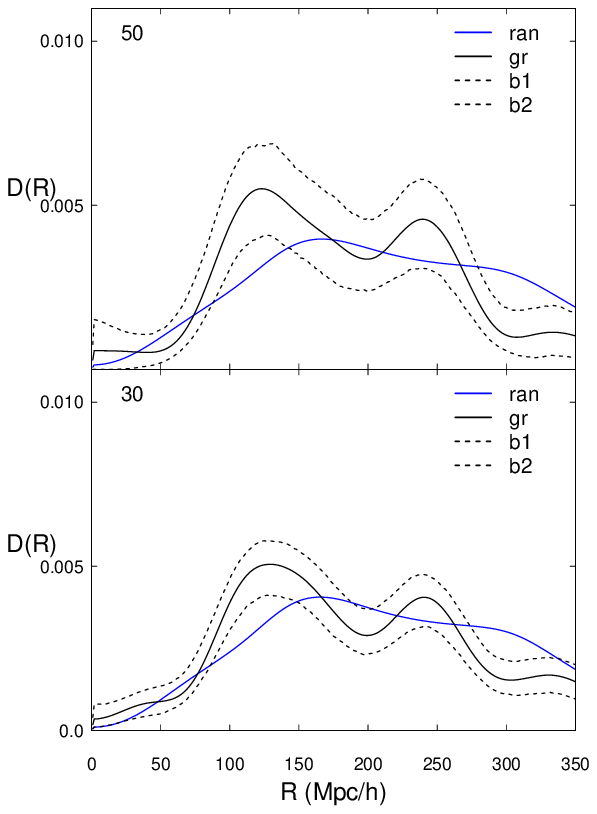}}
\resizebox{0.25\textwidth}{!}{\includegraphics[angle=0]{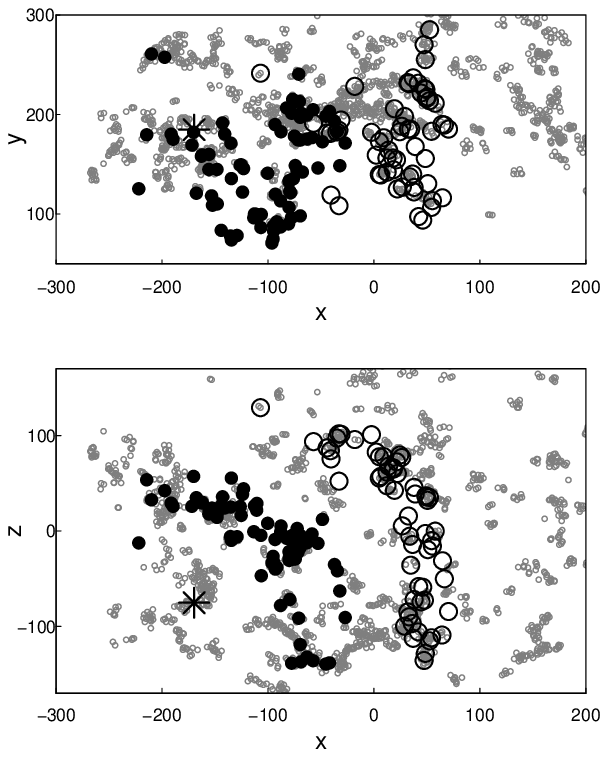}}
\caption{
Distance distribution between cluster Gr34513 (A2020) and other
groups and clusters of different richness (left panels)
and the distribution of groups with at least
four galaxies in Cartesian  coordinates
(right panels). Notations and details are the same as in Fig.~\ref{fig:gr50647}.
The maxima are located at $122$ and $239$~\Mpc\ (upper left panel, 
$N_{\mathrm{gal}} \geq 50$), and $128$ and $240$~\Mpc\ (lower left panel, 
$N_{\mathrm{gal}} \geq 30$). 
} 
\label{fig:gr34513}
\end{figure}

Cluster A2142 (Gr29587), the central cluster of a rich supercluster
(A2142 supercluster, E2015)
is the centre of a possible 
shell whose walls are formed by superclusters  SCl~007, SCl~352, and SCl~349 
(the centre of another shell candidate)
with a density contrast $D8 = 8.5 - 17.0$ (Fig.~\ref{fig:gr29587}).  
The Sloan Great Wall and other rich superclusters
cause a maximum in the cluster distance distribution
at  $D > $ 200~\Mpc\ (Fig.~\ref{fig:gr29587} left panel). 
For this cluster the maxima test showed, however, that the maximum at $135$~\Mpc\
can also be found in the
random distributions. The low significance of the maximum may arise from 
the location of this cluster near the sample edge.

Cluster Gr20159 (A1026) is an isolated cluster at the edge of a chain of the
Ursa Major (SCl~211) and other superclusters (Fig.~\ref{fig:gr20159}). 
The possible shell around it is formed by
poor superclusters and clusters at the edges of voids.  
The maximum at $131$~\Mpc\ differs from random in the distance distribution
of rich clusters around it, with $N_{\mathrm{gal}} \geq 50$, and is not
significant in the case of poorer groups. 
This is the only cluster in our sample for which the second maximum in the 
distance distribution can be found from random distributions with  a
probability lower than 0.001. 
For this cluster the second maximum for observed
clusters is strong, the random distribution
is much lower at large distances, but stronger than the observed distribution
at smaller distances due to normalisation.

Cluster Gr3714 (A1139) in Fig.~\ref{fig:gr3714} is located at the edge of a chain of 
galaxy groups connected by rich superclusters with poor groups
through the void. The possible shell around it is formed by
the Bootes supercluster, the Ursa Major, and other superclusters.

Cluster Gr34513 (A2020) is the only rich  cluster in the supercluster
SCl~1311 with a density contrast $D8 = 5.4$, the lowest in our supercluster sample.
However, the possible shell walls around it are formed by very rich
and high-density superclusters, the Sloan Great Wall, and the Bootes supercluster among them
(see Fig.~\ref{fig:gr34513}).

For these two clusters the 
maxima test confirmed that the maximum in the distance distributions
to other groups and clusters at $130$~\Mpc\
has a probability lower than 0.001 to be random.

\section{Discussion and conclusions}
\label{sect:discussion} 

We analysed the distance distributions around rich clusters of galaxies
in the local Universe and found 28 clusters with a maximum
at $120 - 130$~\Mpc\ scales in these distributions. 
These clusters represent centres of six  shell-like structures.

\begin{itemize}
\item{}
The rich cluster A1795 in the Bootes
supercluster has an almost complete shell-like 
structure with a radius of about $120$~\Mpc\ around it.
The probability that we derive this maximum from random
distributions is lower than 0.001.
The walls of the shell are delineated
by filaments of galaxy systems of various richness from
poor groups to the richest superclusters in the nearby Universe -- 
the Sloan Great Wall, the Corona Borealis supercluster, the Ursa Major
supercluster, and others. 

\item{}
We found five additional possible centres
of shell-like systems with a maximum on a scale of about 
$120 - 130$~\Mpc\ 
in the distance distribution of galaxy groups and clusters around them. 
These possible shells are only partly embedded in our sample volume.  
For one cluster (A1026) the maximum at a distance $d > 200$~\Mpc\
cannot be drawn from a random distribution.

\item{}
The luminosity density contrast of possible shell centres and walls varies from the
highest density peak in the whole SDSS survey (in the region of the A2142 supercluster,
$D8 > 20$) 
to relatively low density regions with $D8 < 4$. 

\end{itemize}

The radii of the shells
are larger than $\approx 109$~\Mpc, the well-known BAO scale,
and they are wide, being in the interval of $119 - 135$~\Mpc.
They depend on the positions of the central clusters in the cosmic web. 

Another important difference is the composition of the shells: 
BAO shells are barely detected in the distribution of galaxies. 
In contrast, the walls of the shell connected with cluster 
A1795 are formed by superclusters,  
structures of the highest density in the nearby Universe. 

These differences in the properties of shells  
suggest that the causes for them are different phenomena.

A characteristic scale of
about $120 - 130$~\Mpc\ has been found in the
distribution of rich superclusters 
\citep{1994MNRAS.269..301E, 1997Natur.385..139E, 2002A&A...393....1S}, and in the correlation
function of galaxy clusters \citep{1997MNRAS.289..801E, 1997MNRAS.289..813E}. 
This is a manifestation  of the cellular structure of the Universe,
delineated by rich clusters and galaxy superclusters
\citep{1977TarOP...1A...1J, 1978IAUS...79..241J, 2010MNRAS.408.2163A, 2014arXiv1409.8661A}.
\citet{1997A&AS..123..119E} noted 
that while the characteristic sizes
of voids between superclusters of galaxies are similar for
superclusters of all richnesses, the supercluster richness differs
in wall voids. This
agrees with our present finding that the richness of structures and their luminosity
density  in shell centres and walls varies strongly.

The pattern of the cosmic web originates from processes in the early universe.  
The positions of high-density
peaks and voids do not change much during the cosmic evolution, 
only their amplitude 
grows with time (see references in Introduction).
However, it is not yet clear which processes cause shell-like 
structures in the local cosmic web.
Therefore it is important
to continue the analysis of the possible characteristic scales in
the galaxy cluster and supercluster distribution using  samples
with larger sky coverage and wider redshift range
to understand the presence and possible nature of these scales.

\begin{acknowledgements}

We cordially thank the referee for many useful comments and suggestions
that greatly helped us to improve the paper.

We are pleased to thank the SDSS Team for the publicly available data
releases.  
Funding for the Sloan Digital Sky Survey (SDSS) and SDSS-II has been
  provided by the Alfred P. Sloan Foundation, the Participating Institutions,
  the National Science Foundation, the U.S.  Department of Energy, the
  National Aeronautics and Space Administration, the Japanese Monbukagakusho,
  and the Max Planck Society, and the Higher Education Funding Council for
  England.  The SDSS Web site is \texttt{http://www.sdss.org/}.
  The SDSS is managed by the Astrophysical Research Consortium (ARC) for the
  Participating Institutions.  The Participating Institutions are the American
  Museum of Natural History, Astrophysical Institute Potsdam, University of
  Basel, University of Cambridge, Case Western Reserve University, The
  University of Chicago, Drexel University, Fermilab, the Institute for
  Advanced Study, the Japan Participation Group, The Johns Hopkins University,
  the Joint Institute for Nuclear Astrophysics, the Kavli Institute for
  Particle Astrophysics and Cosmology, the Korean Scientist Group, the Chinese
  Academy of Sciences (LAMOST), Los Alamos National Laboratory, the
  Max-Planck-Institute for Astronomy (MPIA), the Max-Planck-Institute for
  Astrophysics (MPA), New Mexico State University, Ohio State University,
  University of Pittsburgh, University of Portsmouth, Princeton University,
  the United States Naval Observatory, and the University of Washington.

The present study was supported by  by the ETAG projects 
IUT26-2  and IUT40-2, and by the European Structural Funds
grant for the Centre of Excellence "Dark Matter in (Astro)particle Physics and
Cosmology" TK120. This work has also been supported by
ICRAnet through a professorship for Jaan Einasto.
VJM, PAM, and LlHG acknowledge support from the Spanish Ministry for Economy and
Competitiveness and FEDER funds through grants AYA2010-22111-C03-02 and
AYA2013-48623-C2-2, and Generalitat Valenciana project PrometeoII 2014/060.

\end{acknowledgements}

\bibliographystyle{aa}
\bibliography{shell.bib}

\begin{thebibliography}{55}
\expandafter\ifx\csname natexlab\endcsname\relax\def\natexlab#1{#1}\fi

\bibitem[{{Aihara} {et~al.}(2011){Aihara}, {Allende Prieto}, {An}, {Anderson},
  {Aubourg}, {Balbinot}, {Beers}, {Berlind}, {Bickerton}, {Bizyaev}, {Blanton},
  {Bochanski}, {Bolton}, {Bovy}, {Brandt}, {Brinkmann}, {Brown}, {Brownstein},
  {Busca}, {Campbell}, {Carr}, {Chen}, {Chiappini}, {Comparat}, {Connolly},
  {Cortes}, {Croft}, {Cuesta}, {da Costa}, {Davenport}, {Dawson}, {Dhital},
  {Ealet}, {Ebelke}, {Edmondson}, {Eisenstein}, {Escoffier}, {Esposito},
  {Evans}, {Fan}, {Femen{\'{\i}}a Castell{\'a}}, {Font-Ribera}, {Frinchaboy},
  {Ge}, {Gillespie}, {Gilmore}, {Gonz{\'a}lez Hern{\'a}ndez}, {Gott}, {Gould},
  {Grebel}, {Gunn}, {Hamilton}, {Harding}, {Harris}, {Hawley}, {Hearty}, {Ho},
  {Hogg}, {Holtzman}, {Honscheid}, {Inada}, {Ivans}, {Jiang}, {Johnson},
  {Jordan}, {Jordan}, {Kazin}, {Kirkby}, {Klaene}, {Knapp}, {Kneib},
  {Kochanek}, {Koesterke}, {Kollmeier}, {Kron}, {Lampeitl}, {Lang}, {Le Goff},
  {Lee}, {Lin}, {Long}, {Loomis}, {Lucatello}, {Lundgren}, {Lupton}, {Ma},
  {MacDonald}, {Mahadevan}, {Maia}, {Makler}, {Malanushenko}, {Malanushenko},
  {Mandelbaum}, {Maraston}, {Margala}, {Masters}, {McBride}, {McGehee},
  {McGreer}, {M{\'e}nard}, {Miralda-Escud{\'e}}, {Morrison}, {Mullally},
  {Muna}, {Munn}, {Murayama}, {Myers}, {Naugle}, {Fausti Neto}, {Cuong Nguyen},
  {Nichol}, {O'Connell}, {Ogando}, {Olmstead}, {Oravetz}, {Padmanabhan},
  {Palanque-Delabrouille}, {Pan}, {Pandey}, {P{\^a}ris}, {Percival},
  {Petitjean}, {Pfaffenberger}, {Pforr}, {Phleps}, {Pichon}, {Pieri}, {Prada},
  {Price-Whelan}, {Raddick}, {Ramos}, {Reyl{\'e}}, {Rich}, {Richards}, {Rix},
  {Robin}, {Rocha-Pinto}, {Rockosi}, {Roe}, {Rollinde}, {Ross}, {Ross},
  {Rossetto}, {S{\'a}nchez}, {Sayres}, {Schlegel}, {Schlesinger}, {Schmidt},
  {Schneider}, {Sheldon}, {Shu}, {Simmerer}, {Simmons}, {Sivarani}, {Snedden},
  {Sobeck}, {Steinmetz}, {Strauss}, {Szalay}, {Tanaka}, {Thakar}, {Thomas},
  {Tinker}, {Tofflemire}, {Tojeiro}, {Tremonti}, {Vandenberg}, {Vargas
  Maga{\~n}a}, {Verde}, {Vogt}, {Wake}, {Wang}, {Weaver}, {Weinberg}, {White},
  {White}, {Yanny}, {Yasuda}, {Yeche}, \& {Zehavi}}]{2011ApJS..193...29A}
{Aihara}, H., {Allende Prieto}, C., {An}, D., {et~al.} 2011, \apjs, 193, 29

\bibitem[{{Anderson} {et~al.}(2012){Anderson}, {Aubourg}, {Bailey}, {Bizyaev},
  {Blanton}, {Bolton}, {Brinkmann}, {Brownstein}, {Burden}, {Cuesta}, {da
  Costa}, {Dawson}, {de Putter}, {Eisenstein}, {Gunn}, {Guo}, {Hamilton},
  {Harding}, {Ho}, {Honscheid}, {Kazin}, {Kirkby}, {Kneib}, {Labatie},
  {Loomis}, {Lupton}, {Malanushenko}, {Malanushenko}, {Mandelbaum}, {Manera},
  {Maraston}, {McBride}, {Mehta}, {Mena}, {Montesano}, {Muna}, {Nichol},
  {Nuza}, {Olmstead}, {Oravetz}, {Padmanabhan}, {Palanque-Delabrouille}, {Pan},
  {Parejko}, {P{\^a}ris}, {Percival}, {Petitjean}, {Prada}, {Reid}, {Roe},
  {Ross}, {Ross}, {Samushia}, {S{\'a}nchez}, {Schlegel}, {Schneider},
  {Sc{\'o}ccola}, {Seo}, {Sheldon}, {Simmons}, {Skibba}, {Strauss}, {Swanson},
  {Thomas}, {Tinker}, {Tojeiro}, {Maga{\~n}a}, {Verde}, {Wagner}, {Wake},
  {Weaver}, {Weinberg}, {White}, {Xu}, {Y{\`e}che}, {Zehavi}, \&
  {Zhao}}]{2012MNRAS.427.3435A}
{Anderson}, L., {Aubourg}, E., {Bailey}, S., {et~al.} 2012, \mnras, 427, 3435

\bibitem[{{Arag{\'o}n-Calvo}(2014)}]{2014arXiv1409.8661A}
{Arag{\'o}n-Calvo}, M.~A. 2014, ArXiv e-print [\eprint[arXiv]{1409.8661}]

\bibitem[{{Arag{\'o}n-Calvo} {et~al.}(2010){Arag{\'o}n-Calvo}, {van de
  Weygaert}, \& {Jones}}]{2010MNRAS.408.2163A}
{Arag{\'o}n-Calvo}, M.~A., {van de Weygaert}, R., \& {Jones}, B.~J.~T. 2010,
  \mnras, 408, 2163

\bibitem[{{Arnalte-Mur} {et~al.}(2012){Arnalte-Mur}, {Labatie}, {Clerc},
  {Mart{\'{\i}}nez}, {Starck}, {Lachi{\`e}ze-Rey}, {Saar}, \&
  {Paredes}}]{2012A&A...542A..34A}
{Arnalte-Mur}, P., {Labatie}, A., {Clerc}, N., {et~al.} 2012, \aap, 542, A34

\bibitem[{{Beutler} {et~al.}(2011){Beutler}, {Blake}, {Colless}, {Jones},
  {Staveley-Smith}, {Campbell}, {Parker}, {Saunders}, \&
  {Watson}}]{2011MNRAS.416.3017B}
{Beutler}, F., {Blake}, C., {Colless}, M., {et~al.} 2011, \mnras, 416, 3017

\bibitem[{{Blake} {et~al.}(2011){Blake}, {Kazin}, {Beutler}, {Davis},
  {Parkinson}, {Brough}, {Colless}, {Contreras}, {Couch}, {Croom}, {Croton},
  {Drinkwater}, {Forster}, {Gilbank}, {Gladders}, {Glazebrook}, {Jelliffe},
  {Jurek}, {Li}, {Madore}, {Martin}, {Pimbblet}, {Poole}, {Pracy}, {Sharp},
  {Wisnioski}, {Woods}, {Wyder}, \& {Yee}}]{2011MNRAS.418.1707B}
{Blake}, C., {Kazin}, E.~A., {Beutler}, F., {et~al.} 2011, \mnras, 418, 1707

\bibitem[{{Blanton} {et~al.}(2003){Blanton}, {Hogg}, {Bahcall}, {Brinkmann},
  {Britton}, {Connolly}, {Csabai}, {Fukugita}, {Loveday}, {Meiksin}, {Munn},
  {Nichol}, {Okamura}, {Quinn}, {Schneider}, {Shimasaku}, {Strauss}, {Tegmark},
  {Vogeley}, \& {Weinberg}}]{blanton03b}
{Blanton}, M.~R., {Hogg}, D.~W., {Bahcall}, N.~A., {et~al.} 2003, \apj, 592,
  819

\bibitem[{{Blanton} \& {Roweis}(2007)}]{2007AJ....133..734B}
{Blanton}, M.~R. \& {Roweis}, S. 2007, \aj, 133, 734

\bibitem[{{Bond} {et~al.}(1996){Bond}, {Kofman}, \&
  {Pogosyan}}]{1996Natur.380..603B}
{Bond}, J.~R., {Kofman}, L., \& {Pogosyan}, D. 1996, \nat, 380, 603

\bibitem[{{Chincarini} \& {Rood}(1979)}]{1979ApJ...230..648C}
{Chincarini}, G. \& {Rood}, H.~J. 1979, \apj, 230, 648

\bibitem[{{Cole} {et~al.}(2005){Cole}, {Percival}, {Peacock}, {Norberg},
  {Baugh}, {Frenk}, {Baldry}, {Bland-Hawthorn}, {Bridges}, {Cannon}, {Colless},
  {Collins}, {Couch}, {Cross}, {Dalton}, {Eke}, {De Propris}, {Driver},
  {Efstathiou}, {Ellis}, {Glazebrook}, {Jackson}, {Jenkins}, {Lahav}, {Lewis},
  {Lumsden}, {Maddox}, {Madgwick}, {Peterson}, {Sutherland}, \&
  {Taylor}}]{2005MNRAS.362..505C}
{Cole}, S., {Percival}, W.~J., {Peacock}, J.~A., {et~al.} 2005, \mnras, 362,
  505

\bibitem[{{Davison} \& {Hinkley}(1997)}]{davhink97}
{Davison}, A.~C. \& {Hinkley}, D.~V. 1997, {Bootstrap Methods and their
  Application} (Cambridge University Press, Cambridge)

\bibitem[{{de Lapparent} {et~al.}(1986){de Lapparent}, {Geller}, \&
  {Huchra}}]{1986ApJ...302L...1D}
{de Lapparent}, V., {Geller}, M.~J., \& {Huchra}, J.~P. 1986, \apjl, 302, L1

\bibitem[{{Einasto} {et~al.}(1997{\natexlab{a}}){Einasto}, {Einasto}, {Frisch},
  {Gottlober}, {Muller}, {Saar}, {Starobinsky}, {Tago}, {Tucker}, \&
  {Andernach}}]{1997MNRAS.289..801E}
{Einasto}, J., {Einasto}, M., {Frisch}, P., {et~al.} 1997{\natexlab{a}},
  \mnras, 289, 801

\bibitem[{{Einasto} {et~al.}(1997{\natexlab{b}}){Einasto}, {Einasto}, {Frisch},
  {Gottlober}, {Muller}, {Saar}, {Starobinsky}, \&
  {Tucker}}]{1997MNRAS.289..813E}
{Einasto}, J., {Einasto}, M., {Frisch}, P., {et~al.} 1997{\natexlab{b}},
  \mnras, 289, 813

\bibitem[{{Einasto} {et~al.}(1997{\natexlab{c}}){Einasto}, {Einasto},
  {Gottl{\"o}ber}, {M{\"u}ller}, {Saar}, {Starobinsky}, {Tago}, {Tucker},
  {Andernach}, \& {Frisch}}]{1997Natur.385..139E}
{Einasto}, J., {Einasto}, M., {Gottl{\"o}ber}, S., {et~al.} 1997{\natexlab{c}},
  \nat, 385, 139

\bibitem[{{Einasto} {et~al.}(1975){Einasto}, {J{\^o}eveer}, {Kivila}, \&
  {Tago}}]{1975ATsir.895....2E}
{Einasto}, J., {J{\^o}eveer}, M., {Kivila}, A., \& {Tago}, E. 1975,
  Astronomicheskij Tsirkulyar, 895, 2

\bibitem[{{Einasto} {et~al.}(2011{\natexlab{a}}){Einasto}, {Suhhonenko},
  {H{\"u}tsi}, {Saar}, {Einasto}, {Liivam{\"a}gi}, {M{\"u}ller}, {Starobinsky},
  {Tago}, \& {Tempel}}]{2011A&A...534A.128E}
{Einasto}, J., {Suhhonenko}, I., {H{\"u}tsi}, G., {et~al.} 2011{\natexlab{a}},
  \aap, 534, A128

\bibitem[{{Einasto} {et~al.}(1994){Einasto}, {Einasto}, {Tago}, {Dalton}, \&
  {Andernach}}]{1994MNRAS.269..301E}
{Einasto}, M., {Einasto}, J., {Tago}, E., {Dalton}, G.~B., \& {Andernach}, H.
  1994, \mnras, 269, 301

\bibitem[{{Einasto} {et~al.}(2001){Einasto}, {Einasto}, {Tago}, {M{\"u}ller},
  \& {Andernach}}]{2001AJ....122.2222E}
{Einasto}, M., {Einasto}, J., {Tago}, E., {M{\"u}ller}, V., \& {Andernach}, H.
  2001, \aj, 122, 2222

\bibitem[{{Einasto} {et~al.}(2015){Einasto}, {Gramann}, {Saar},
  {Liivam{\"a}gi}, {Tempel}, {Nevalainen}, {Hein{\"a}m{\"a}ki}, {Park}, \&
  {Einasto}}]{2015A&A...580A..69E}
{Einasto}, M., {Gramann}, M., {Saar}, E., {et~al.} 2015, \aap, 580, A69

\bibitem[{{Einasto} {et~al.}(2011{\natexlab{b}}){Einasto}, {Liivam{\"a}gi},
  {Tago}, {Saar}, {Tempel}, {Einasto}, {Mart{\'{\i}}nez}, \&
  {Hein{\"a}m{\"a}ki}}]{2011A&A...532A...5E}
{Einasto}, M., {Liivam{\"a}gi}, L.~J., {Tago}, E., {et~al.} 2011{\natexlab{b}},
  \aap, 532, A5

\bibitem[{{Einasto} {et~al.}(2012{\natexlab{a}}){Einasto}, {Liivam{\"a}gi},
  {Tempel}, {Saar}, {Vennik}, {Nurmi}, {Gramann}, {Einasto}, {Tago},
  {Hein{\"a}m{\"a}ki}, {Ahvensalmi}, \&
  {Mart{\'{\i}}nez}}]{2012A&A...542A..36E}
{Einasto}, M., {Liivam{\"a}gi}, L.~J., {Tempel}, E., {et~al.}
  2012{\natexlab{a}}, \aap, 542, A36

\bibitem[{{Einasto} {et~al.}(1997{\natexlab{d}}){Einasto}, {Tago}, {Jaaniste},
  {Einasto}, \& {Andernach}}]{1997A&AS..123..119E}
{Einasto}, M., {Tago}, E., {Jaaniste}, J., {Einasto}, J., \& {Andernach}, H.
  1997{\natexlab{d}}, \aaps, 123, 119

\bibitem[{{Einasto} {et~al.}(2014){Einasto}, {Tago}, {Lietzen}, {Park},
  {Hein{\"a}m{\"a}ki}, {Saar}, {Song}, {Liivam{\"a}gi}, \&
  {Einasto}}]{2014A&A...568A..46E}
{Einasto}, M., {Tago}, E., {Lietzen}, H., {et~al.} 2014, \aap, 568, A46

\bibitem[{{Einasto} {et~al.}(2012{\natexlab{b}}){Einasto}, {Vennik}, {Nurmi},
  {Tempel}, {Ahvensalmi}, {Tago}, {Liivam{\"a}gi}, {Saar}, {Hein{\"a}m{\"a}ki},
  {Einasto}, \& {Mart{\'{\i}}nez}}]{2012A&A...540A.123E}
{Einasto}, M., {Vennik}, J., {Nurmi}, P., {et~al.} 2012{\natexlab{b}}, \aap,
  540, A123

\bibitem[{{Eisenstein} {et~al.}(2005){Eisenstein}, {Zehavi}, {Hogg},
  {Scoccimarro}, {Blanton}, {Nichol}, {Scranton}, {Seo}, {Tegmark}, {Zheng},
  {Anderson}, {Annis}, {Bahcall}, {Brinkmann}, {Burles}, {Castander},
  {Connolly}, {Csabai}, {Doi}, {Fukugita}, {Frieman}, {Glazebrook}, {Gunn},
  {Hendry}, {Hennessy}, {Ivezi{\'c}}, {Kent}, {Knapp}, {Lin}, {Loh}, {Lupton},
  {Margon}, {McKay}, {Meiksin}, {Munn}, {Pope}, {Richmond}, {Schlegel},
  {Schneider}, {Shimasaku}, {Stoughton}, {Strauss}, {SubbaRao}, {Szalay},
  {Szapudi}, {Tucker}, {Yanny}, \& {York}}]{2005ApJ...633..560E}
{Eisenstein}, D.~J., {Zehavi}, I., {Hogg}, D.~W., {et~al.} 2005, \apj, 633, 560

\bibitem[{{Gregory} \& {Thompson}(1978)}]{1978ApJ...222..784G}
{Gregory}, S.~A. \& {Thompson}, L.~A. 1978, \apj, 222, 784

\bibitem[{{J{\~o}eveer} \& {Einasto}(1978)}]{1978IAUS...79..241J}
{J{\~o}eveer}, M. \& {Einasto}, J. 1978, in IAU Symposium, Vol.~79, Large Scale
  Structures in the Universe, ed. M.~S. {Longair} \& J.~{Einasto}, 241--250

\bibitem[{{J{\~o}eveer} {et~al.}(1977){J{\~o}eveer}, {Einasto}, \&
  {Tago}}]{1977TarOP...1A...1J}
{J{\~o}eveer}, M., {Einasto}, J., \& {Tago}, M. 1977, Tartu
  Astrof{\"u}{\"u}s.~Obs.~Preprint, Nr.~A-1, 45 p., 1, A1

\bibitem[{{Juszkiewicz} {et~al.}(2013){Juszkiewicz}, {Hellwing}, \& {van de
  Weygaert}}]{2013MNRAS.429.1206J}
{Juszkiewicz}, R., {Hellwing}, W.~A., \& {van de Weygaert}, R. 2013, \mnras,
  429, 1206

\bibitem[{{Kirshner} {et~al.}(1981){Kirshner}, {Oemler}, {Schechter}, \&
  {Shectman}}]{1981ApJ...248L..57K}
{Kirshner}, R.~P., {Oemler}, Jr., A., {Schechter}, P.~L., \& {Shectman}, S.~A.
  1981, \apjl, 248, L57

\bibitem[{{Kofman} \& {Shandarin}(1988)}]{1988Natur.334..129K}
{Kofman}, L.~A. \& {Shandarin}, S.~F. 1988, \nat, 334, 129

\bibitem[{{Ledlow} {et~al.}(2003){Ledlow}, {Voges}, {Owen}, \&
  {Burns}}]{2003AJ....126.2740L}
{Ledlow}, M.~J., {Voges}, W., {Owen}, F.~N., \& {Burns}, J.~O. 2003, \aj, 126,
  2740

\bibitem[{{Liivam{\"a}gi} {et~al.}(2012){Liivam{\"a}gi}, {Tempel}, \&
  {Saar}}]{2012A&A...539A..80L}
{Liivam{\"a}gi}, L.~J., {Tempel}, E., \& {Saar}, E. 2012, \aap, 539, A80

\bibitem[{{Mart{\'{\i}}nez} {et~al.}(2009){Mart{\'{\i}}nez}, {Arnalte-Mur},
  {Saar}, {de la Cruz}, {Pons-Border{\'{\i}}a}, {Paredes},
  {Fern{\'a}ndez-Soto}, \& {Tempel}}]{2009ApJ...696L..93M}
{Mart{\'{\i}}nez}, V.~J., {Arnalte-Mur}, P., {Saar}, E., {et~al.} 2009, \apjl,
  696, L93

\bibitem[{{Mart{\'{\i}}nez} \& {Saar}(2002)}]{mar03}
{Mart{\'{\i}}nez}, V.~J. \& {Saar}, E. 2002, {Statistics of the Galaxy
  Distribution} (Chapman {\&} Hall/CRC, Boca Raton)

\bibitem[{{Miller} {et~al.}(2004){Miller}, {Croom}, {Boyle}, {Loaring},
  {Smith}, {Shanks}, \& {Outram}}]{2004MNRAS.355..385M}
{Miller}, L., {Croom}, S.~M., {Boyle}, B.~J., {et~al.} 2004, \mnras, 355, 385

\bibitem[{{Oort}(1983)}]{1983ARA&A..21..373O}
{Oort}, J.~H. 1983, \araa, 21, 373

\bibitem[{{Padmanabhan} {et~al.}(2012){Padmanabhan}, {Xu}, {Eisenstein},
  {Scalzo}, {Cuesta}, {Mehta}, \& {Kazin}}]{2012MNRAS.427.2132P}
{Padmanabhan}, N., {Xu}, X., {Eisenstein}, D.~J., {et~al.} 2012, \mnras, 427,
  2132

\bibitem[{{Park} {et~al.}(2007){Park}, {Choi}, {Vogeley}, {Gott}, \&
  {Blanton}}]{2007ApJ...658..898P}
{Park}, C., {Choi}, Y., {Vogeley}, M.~S., {Gott}, III, J.~R., \& {Blanton},
  M.~R. 2007, \apj, 658, 898

\bibitem[{{Peebles}(1980)}]{1980lssu.book.....P}
{Peebles}, P.~J.~E. 1980, {The large-scale structure of the universe}
  (Princeton University Press)

\bibitem[{{Peebles} \& {Yu}(1970)}]{1970ApJ...162..815P}
{Peebles}, P.~J.~E. \& {Yu}, J.~T. 1970, \apj, 162, 815

\bibitem[{{Planck Collaboration} {et~al.}(2015){Planck Collaboration}, {Ade},
  {Aghanim}, {Arnaud}, {Ashdown}, {Aumont}, {Baccigalupi}, {Banday},
  {Barreiro}, {Bartlett}, \& et~al.}]{2015arXiv150201589P}
{Planck Collaboration}, {Ade}, P.~A.~R., {Aghanim}, N., {et~al.} 2015, ArXiv
  e-print [\eprint[arXiv]{1502.01589}]

\bibitem[{{Saar} {et~al.}(2002){Saar}, {Einasto}, {Toomet}, {Starobinsky},
  {Andernach}, {Einasto}, {Kasak}, \& {Tago}}]{2002A&A...393....1S}
{Saar}, E., {Einasto}, J., {Toomet}, O., {et~al.} 2002, \aap, 393, 1

\bibitem[{{Silverman}(1986)}]{1986desd.book.....S}
{Silverman}, B.~W. 1986, {Density estimation for statistics and data analysis}
  (Chapman and Hall, London)

\bibitem[{{Springel} {et~al.}(2005){Springel}, {White}, {Jenkins}, {Frenk},
  {Yoshida}, {Gao}, {Navarro}, {Thacker}, {Croton}, {Helly}, {Peacock}, {Cole},
  {Thomas}, {Couchman}, {Evrard}, {Colberg}, \& {Pearce}}]{2005Natur.435..629S}
{Springel}, V., {White}, S.~D.~M., {Jenkins}, A., {et~al.} 2005, \nat, 435, 629

\bibitem[{{Suhhonenko} {et~al.}(2011){Suhhonenko}, {Einasto}, {Liivam{\"a}gi},
  {Saar}, {Einasto}, {H{\"u}tsi}, {M{\"u}ller}, {Starobinsky}, {Tago}, \&
  {Tempel}}]{2011A&A...531A.149S}
{Suhhonenko}, I., {Einasto}, J., {Liivam{\"a}gi}, L.~J., {et~al.} 2011, \aap,
  531, A149

\bibitem[{{Tempel} {et~al.}(2012){Tempel}, {Tago}, \&
  {Liivam{\"a}gi}}]{2012A&A...540A.106T}
{Tempel}, E., {Tago}, E., \& {Liivam{\"a}gi}, L.~J. 2012, \aap, 540, A106

\bibitem[{{Tempel} {et~al.}(2014){Tempel}, {Tamm}, {Gramann}, {Tuvikene},
  {Liivam{\"a}gi}, {Suhhonenko}, {Kipper}, {Einasto}, \&
  {Saar}}]{2014A&A...566A...1T}
{Tempel}, E., {Tamm}, A., {Gramann}, M., {et~al.} 2014, \aap, 566, A1

\bibitem[{{Turner} \& {Gott}(1976)}]{tg76}
{Turner}, E.~L. \& {Gott}, III, J.~R. 1976, \apjs, 32, 409

\bibitem[{{van de Weygaert} \& {Schaap}(2009)}]{2009LNP...665..291V}
{van de Weygaert}, R. \& {Schaap}, W. 2009, in Lecture Notes in Physics, Berlin
  Springer Verlag, Vol. 665, Data Analysis in Cosmology, ed. V.~J.
  {Mart{\'{\i}}nez}, E.~{Saar}, E.~{Mart{\'{\i}}nez-Gonz{\'a}lez}, \& M.-J.
  {Pons-Border{\'{\i}}a}, 291--413

\bibitem[{{Weinberg} {et~al.}(2013){Weinberg}, {Mortonson}, {Eisenstein},
  {Hirata}, {Riess}, \& {Rozo}}]{2013PhR...530...87W}
{Weinberg}, D.~H., {Mortonson}, M.~J., {Eisenstein}, D.~J., {et~al.} 2013,
  \physrep, 530, 87

\bibitem[{{Zeldovich} {et~al.}(1982){Zeldovich}, {Einasto}, \&
  {Shandarin}}]{1982Natur.300..407Z}
{Zeldovich}, I.~B., {Einasto}, J., \& {Shandarin}, S.~F. 1982, \nat, 300, 407

\end{thebibliography}

\end{document}